\documentclass{article}

\usepackage{microtype}
\usepackage{graphicx}
\usepackage{subfigure}
\usepackage{booktabs} 

\usepackage{hyperref}

\usepackage[accepted]{icml2025}

\usepackage{amsmath}
\usepackage{amssymb}
\usepackage{mathtools}
\usepackage{amsthm}
\usepackage{graphicx}  
\usepackage{pifont} 
\usepackage{amsmath} 
\usepackage{array}
\usepackage{makecell} 
\usepackage{tabularx}
\usepackage{tcolorbox}
\usepackage[capitalize,noabbrev]{cleveref}

\theoremstyle{plain}

\theoremstyle{definition}

\theoremstyle{remark}

\usepackage[textsize=tiny]{todonotes}

\icmltitlerunning{DisProtEdit: Exploring Disentangled Representations for Multi-Attribute Protein Editing}

\begin{document}

\twocolumn[
\icmltitle{DisProtEdit: Exploring Disentangled Representations for Multi-Attribute Protein Editing}



\icmlsetsymbol{equal}{*}

\begin{icmlauthorlist}
\icmlauthor{Max Ku}{uwat}
\icmlauthor{Sun Sun}{uwat,nrc}
\icmlauthor{Hongyu Guo}{nrc}
\icmlauthor{Wenhu Chen}{uwat}
\end{icmlauthorlist}

\icmlaffiliation{nrc}{National Research Council Canada, Ottawa, Canada}
\icmlaffiliation{uwat}{Cheriton School of Computer Science, University of Waterloo, Waterloo, Canada}

\icmlcorrespondingauthor{Max Ku}{m3ku@uwaterloo.ca}
\icmlcorrespondingauthor{Wenhu Chen}{wenhuchen@uwaterloo.ca}

\icmlkeywords{protein editing, disentanglement}

\centering{\url{https://tiger-ai-lab.github.io/DisProtEdit/}}
\vskip 0.2in
]



\printAffiliationsAndNotice{}  

\begin{abstract}
We introduce DisProtEdit, a controllable protein editing framework that leverages dual-channel natural language supervision to learn disentangled representations of structural and functional properties. 
Unlike prior approaches that rely on joint holistic embeddings, DisProtEdit explicitly separates semantic factors, enabling modular and interpretable control. To support this, we construct SwissProtDis, a large-scale multimodal dataset where each protein sequence is paired with two textual descriptions, one for structure and one for function, automatically decomposed using a large language model. DisProtEdit aligns protein and text embeddings using alignment and uniformity objectives, while a disentanglement loss promotes independence between structural and functional semantics. At inference time, protein editing is performed by modifying one or both text inputs and decoding from the updated latent representation. Experiments on protein editing and representation learning benchmarks demonstrate that DisProtEdit performs competitively with existing methods while providing improved interpretability and controllability. On a newly constructed multi-attribute editing benchmark, the model achieves a both-hit success rate of up to 61.7\%, highlighting its effectiveness in coordinating simultaneous structural and functional edits.
\end{abstract}

\begin{figure*}[!ht]
    \centering
    \includegraphics[width=0.99\linewidth]{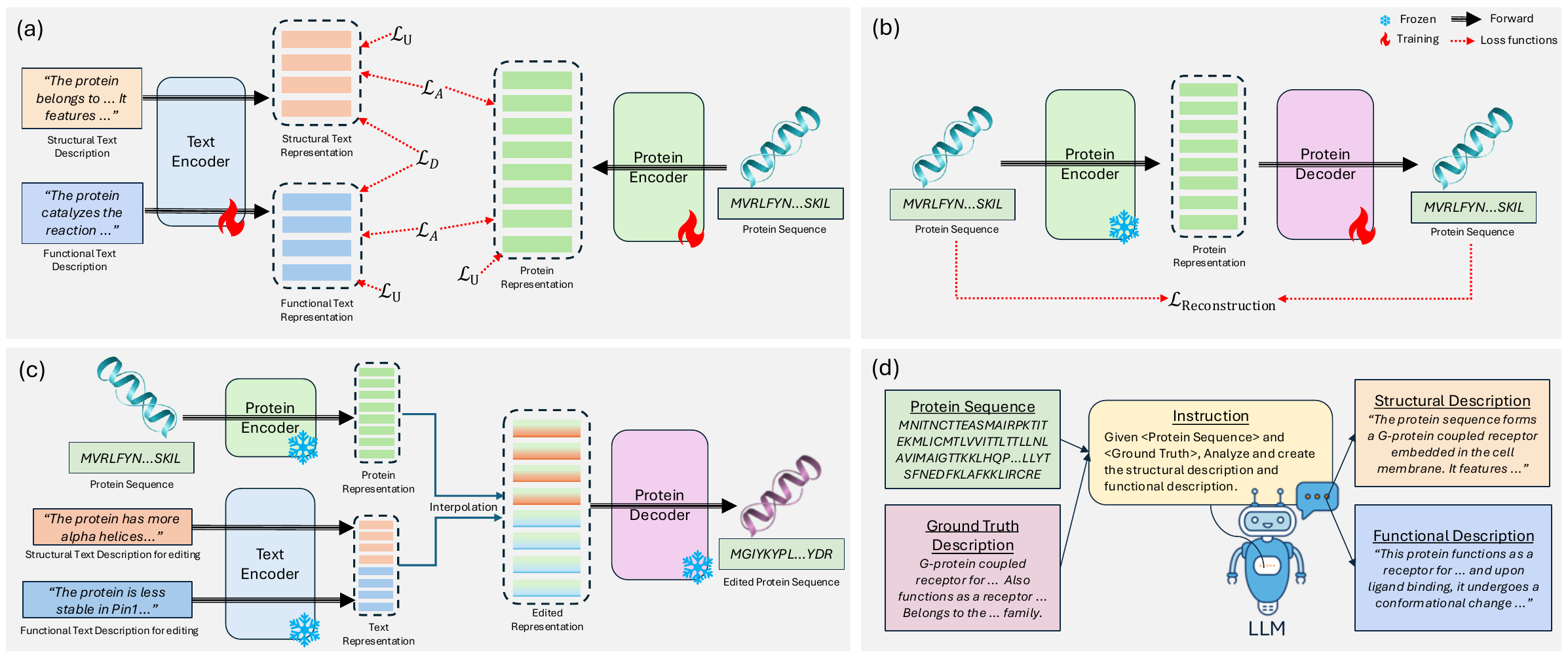}
    \vspace{-1em}
    \caption{
Overview of the DisProtEdit framework. (a) During joint training, proteins and their corresponding structural and functional text descriptions are encoded into a modality-aligned embedding space using alignment, uniformity, and disentanglement objectives. (b) A decoder is trained to reconstruct sequences from latent representations. (c) Protein editing is performed via interpolation between the original embedding and a text-guided embedding. (d) SwissProtDis dataset construction process. Raw annotations are decomposed into structural and functional text descriptions using a large language model.
}
    \label{fig:full-framework}
\end{figure*}

\section{Introduction}


Designing and modeling proteins is a central pursuit in computational biology, with far-reaching implications for understanding disease mechanisms, developing therapeutics, and advancing synthetic biology. Recent breakthroughs in protein foundation models~\cite{AlphaFold2021, esmrives2019biological} have led to remarkable success in tasks such as structure prediction~\cite{Abramson2024} and de novo sequence generation~\cite{Watson2023}. However, despite this progress, editing proteins in a controllable and interpretable manner remains a critical and underexplored challenge. Unlike generation or classification tasks, protein editing requires the fine-grained manipulation of individual semantic properties, such as modifying a protein’s function while preserving its structural fold. Existing models struggle in this setting for two primary reasons: (1) most models treat protein sequences as holistic inputs, without distinguishing between semantically distinct properties such as structure and function. (2) They lack explicit mechanisms for localized control, making it difficult to isolate and edit specific biological attributes without unintended side effects. To enable practical protein editing, we argue that two capabilities are essential: semantic disentanglement of structure and function, and modular control over each. A well-designed editing system should allow precise manipulation of one property (e.g., function) while preserving others (e.g., structure), guided by interpretable conditioning inputs such as natural language descriptions. The ideal goal is to modify only what is intended and leaving the rest unchanged.

To tackle this challenge, we introduce DisProtEdit, a novel framework for controllable protein editing that learns disentangled representations of structural and functional properties through dual-channel natural language supervision. The core idea is to associate each protein sequence with two independently derived textual descriptions: one capturing structural characteristics and the other describing biological function. Our framework employs alignment and uniformity objectives to align text and protein modalities, and incorporates a disentanglement loss based on maximum mean discrepancy (MMD) to ensure that structural and functional semantics remain distinct. To support this paradigm, we construct a new dataset, \textbf{SwissProtDis}, comprising approximately 540,000 protein sequences annotated with these dual-channel descriptions. These descriptions are extracted from an existing protein–text pair dataset SwissProt~\cite{uniprot} and automatically decomposed into structural and functional components using a large language model~\cite{openai2023gpt4}. During training, the model learns to align each protein sequence with both types of textual embeddings. At inference time, editing is performed through the text interface by modifying either the structural or functional description, or both in a compositional manner. This enables semantically grounded modifications to protein representations and yields a representation space well-suited for downstream tasks such as property prediction. Unlike prior multimodal frameworks that rely on contrastive objectives to learn joint embeddings~\cite{liu2023proteinDT}, DisProtEdit learns a shared representation space through alignment and uniformity losses, while explicitly disentangling structure and function using an MMD-based objective. Through extensive evaluations on protein editing and property prediction tasks, we demonstrate that DisProtEdit achieves strong performance while offering interpretability and controllability beyond what existing approaches provide.

To summarize, our contributions are three-fold: (1) We propose \textbf{DisProtEdit}, a framework for protein editing that learns disentangled structural and functional representations from textual descriptions. Our training objective encourages semantic modularity across modalities, enabling effective and interpretable editing.
(2) We introduce \textbf{SwissProtDis}, a large-scale dataset of protein sequences paired with decomposed structural and functional descriptions, automatically derived from UniProt annotations using a large language model.
(3) We demonstrate that DisProtEdit supports both single-attribute and multi-attribute editing, outperforming contrastive baselines and achieving competitive performance on property prediction tasks.

\section{Preliminaries in Representation Learning}

Our framework builds upon several foundational objectives in representation learning, particularly those designed for feature alignment and disentangled representation learning. In this section, we review the key mathematical formulations that serve as building blocks for modern representation learning methods, especially in multi-modal contexts. 

Let $\{x_a^{(i)}\}_{i=1}^{N}$ and $\{x_b^{(j)}\}_{j=1}^{M}$ be two sets of samples drawn from distributions $X_a$ and $X_b$, respectively. These samples can be organized into a paired dataset $\mathcal{D} = \{(x_a^{(i)}, x_b^{(i)})\}_{i=1}^N$, where each pair $(x_a^{(i)}, x_b^{(i)})$ consists of semantically aligned inputs from either the same or different modalities. Corresponding encoders $f_a(\cdot)$ and $f_b(\cdot)$ map the raw inputs to their latent representations, such that $z_a = f_a(x_a)$ and $z_b = f_b(x_b)$ denote the embeddings for $x_a$ and $x_b$, respectively.

\label{sec:contrastive_learning} \textbf{Contrastive Learning.}  When learning from multiple modalities, it is crucial to ensure that the embeddings of paired inputs are close in a shared latent space, while maintaining sufficient diversity across the entire embedding space to avoid collapse. These objectives have been formalized in contrastive learning frameworks~\cite{ oord2019representationlearningcontrastivepredictive, park2020contrastivelearningunpairedimagetoimage, liu2022pretraining, ModalityGap}. A typical contrastive loss for a positive pair $(x_a^{(i)}, x_b^{(i)})$ is defined in Equation~\ref{eq:pre_contrastive} where $\operatorname{sim}(\cdot, \cdot)$ denotes cosine similarity and $\tau > 0$ is a temperature hyperparameter. 

Contrastive learning encourages semantically similar pairs to be close in the latent space while pushing dissimilar pairs apart, thereby improving the discriminability of learned representations~\cite{chen2020simpleframeworkcontrastivelearning}.
\begin{equation}
   \mathcal{L}_{\text{con}} = -\log \frac{\exp \left( \operatorname{sim}(z_a^{(i)}, z_b^{(i)}) / \tau \right)}
   {\sum\limits_{j=1}^N \exp \left( \operatorname{sim}(z_a^{(i)}, z_b^{(j)}) / \tau \right)}
   \label{eq:pre_contrastive}
\end{equation}

\label{sec:alignment_uniformity} 
\textbf{Alignment and Uniformity.} While contrastive learning is effective, it typically requires a large number of negative samples and careful batch design to prevent false negatives, especially in multi-modal settings~\cite{robinson2021contrastivelearninghardnegative}. Here we discuss the alignment and uniformity objectives as a more interpretable version of contrastive learning, and have been shown to achieve comparable or better downstream task performances~\cite{wang2020hypersphere}. Together, these two objectives approximate contrastive learning in the limit of infinitely many negative samples.

\begin{align}
\mathcal{L}_{\text{align}} &= \mathbb{E}_{(x_a, x_b) \sim D} \left\| f_a(x_a^{(i)}) - f_b(x_b^{(i)}) \right\|_2^2
\label{eq:pre_align} \\
\mathcal{L}_{\text{uniform}} &= \log \mathbb{E}_{x \neq x'} \left[ e^{-2 \left\| f(x) - f(x') \right\|_2^2} \right].
\label{eq:pre_uniform}
\end{align}

The alignment loss encourages matching representations to be close, as defined in Equation~\ref{eq:pre_align}. To complement alignment, the uniformity loss promotes dispersion by penalizing embeddings that cluster too tightly. It is defined in Equation~\ref{eq:pre_uniform}, where $f(x)$ denotes any embedding in the batch, from either modality. This loss encourages embeddings to be uniformly distributed on the hypersphere, helping improving generalization.

\label{sec:mmd} \textbf{Maximum Mean Discrepancy (MMD).} While alignment and uniformity focus on pairwise relationships and overall feature dispersion, MMD offers a complementary perspective by comparing entire feature distributions. In our setting, we apply MMD between learned structural and functional embeddings and distinct, angularly-separated priors. This encourages each embedding to occupy an independent subspace, implicitly promoting semantic disentanglement. Similar strategies are common in disentangled representation learning~\cite{mathieu2019disentanglingdisentanglementvariationalautoencoders, louizos2017causaleffectinferencedeep}. MMD is a kernel-based statistical distance used to compare two probability distributions. Unlike Kullback-Leibler divergence or Jensen-Shannon divergence, MMD makes no parametric assumptions and is computed directly from samples~\cite{mmd2012}. The empirical MMD is defined as Equation~\ref{eq:pre_mmd}, where $k(\cdot, \cdot)$ is a positive-definite kernel, commonly the Radial Basis Function (RBF) kernel as $k(x, y) = \exp({-\frac{\|x - y\|^2}{2\sigma^2}})$~\cite{Bro88}. MMD is zero if and only if $X_a = X_b$ when using a characteristic kernel.

\begin{align}
\text{MMD}(X_a, X_b) = &\ \frac{1}{N^2} \sum_{i, i'} k(x_a^{(i)}, x_a^{(i')}) \nonumber \\
& + \frac{1}{M^2} \sum_{j, j'} k(x_b^{(j)}, x_b^{(j')}) \nonumber \\
& - \frac{2}{NM} \sum_{i, j} k(x_a^{(i)}, x_b^{(j)}).
\label{eq:pre_mmd}
\end{align}

Generally, MMD is widely used in domain adaptation~\cite{yan2017mindclassweightbias}, generative modeling~\cite{louizos2017variationalfairautoencoder}, and disentangled representation learning to encourage separation between independent factors~\cite{mathieu2019disentanglingdisentanglementvariationalautoencoders}.

\begin{figure}[t]
    \centering
    \includegraphics[width=1.0\linewidth]{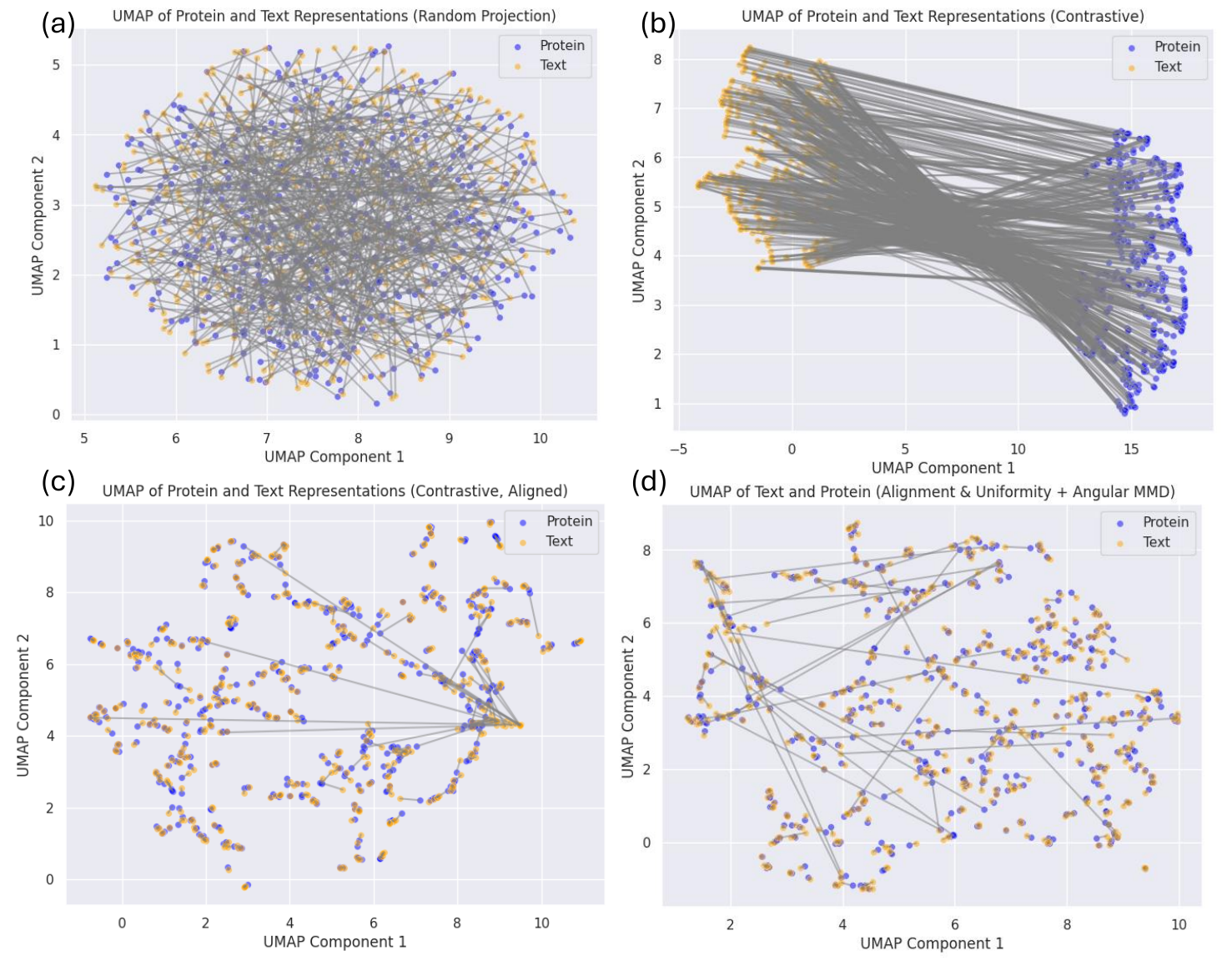}
    \vspace{-1em}
    \caption{
    \textbf{UMAP visualizations of text and protein embeddings under different training strategies.} Each visualization sampled 500 pairs of data. Each point represents a text (yellow) or protein (blue) embedding. Lines connect paired structural/functional text embeddings and their corresponding protein embedding, illustrating the degree of cross-modal alignment. (a) Random projection baseline. (b) Contrastive learning shows a modality gap. (c) Contrastive learning followed by fine-tuning with alignment loss, as common practices in prior works for downstream tasks. (d) Our DisProtEdit framework with alignment, uniformity, and disentanglement objectives. Notably, DisProtEdit achieves alignment quality comparable to (c) without requiring multi-stage training.
    }
    \label{fig:constrastive_comp}
\end{figure}

\section{Method}

\subsection{DisProtEdit Framework}

While structure and function are tightly coupled in biology, recent studies have identified cases of functional divergence among structurally similar proteins~\cite{li2023disentangledwassersteinautoencodertcell}. Inspired by these findings, our approach does not assume complete biological independence, but instead aims to learn controllable axes that enable targeted editing along structure- or function-specific directions. To achieve this, we adopt a disentangled editing framework by aligning protein sequences with dual-channel textual descriptions. Given a protein sequence $x_p$, along with structural and functional descriptions $x_{ts}$ and $x_{tf}$, we train three encoders to map each input into a shared latent space: $z_p = E_p(x_p)$, $z_{ts} = E_{ts}(x_{ts})$, and $z_{tf} = E_{tf}(x_{tf})$, where $E_p(\cdot)$, $E_{ts}(\cdot)$, and $E_{tf}(\cdot)$ denote the protein, structural text, and functional text encoders, respectively. Once trained, we use a protein decoder to reconstruct sequences from the latent representation, enabling controlled editing by manipulating $z_{ts}$ and $z_{tf}$ independently. The full pipeline is illustrated in Figure~\ref{fig:full-framework}.

\textbf{Multimodal Alignment and Uniformity.} To bridge protein and language modalities, we adopt cross-modal alignment and uniformity objectives~\cite{wang2020hypersphere}, which encourage paired embeddings to be close in a shared space while maintaining separation between unrelated samples. Compared to traditional contrastive learning methods~\cite{oord2019representationlearningcontrastivepredictive,chen2020simpleframeworkcontrastivelearning}, this approach improves interpretability and training stability. As discussed in the preliminaries, alignment and uniformity reformulate contrastive learning into two modular objectives: one promoting local similarity and the other encouraging global dispersion, approximately equivariant to contrastive learning in the limit of infinitely many negative samples. This formulation avoids the reliance on large batch sizes and negative sampling, which is particularly beneficial in multimodal settings where negative sampling can introduce false negatives~\cite{robinson2021contrastivelearninghardnegative, huynh2022boostingcontrastiveselfsupervisedlearning}. 

The alignment objective in our context is defined in Equation~\ref{eq:method_align}, where we minimize the distance between the protein embedding and the concatenated structural and functional text embeddings. This encourages consistency across modalities while preserving semantic modularity. Then to prevent representational collapse and improve generalization, we apply uniformity loss that regularizes the geometry of the embedding space as denoted in Equation~\ref{eq:method_uniform} where text embedding $z_t^{(i)}$ is obtained by concatenating $z_{ts}^{(i)}$ and $z_{tf}^{(i)}$. 

\begin{align}
\mathcal{L}_{\text{A}} &= \frac{1}{N} \sum_{i=1}^{N} \left\| \text{concat}(z_{ts}^{(i)}, z_{tf}^{(i)}) - z_{p}^{(i)} \right\|^2
\label{eq:method_align} \\
\mathcal{L}_{\text{U}} &= 
\log \left( \frac{1}{N(N-1)} \sum_{i \ne j} e^{-t \| z_t^{(i)} - z_t^{(j)} \|^2} \right) \nonumber \\
&\quad + \log \left( \frac{1}{N(N-1)} \sum_{i \ne j} e^{-t \| z_p^{(i)} - z_p^{(j)} \|^2} \right).
\label{eq:method_uniform}
\end{align}

This objective penalizes highly concentrated representations by encouraging the embeddings to be uniformly distributed on the hypersphere. In ideal scenario, the concatenated text embedding closely approximates the protein embedding (i.e. $z_t \approx z_p$), reflecting strong cross-modal alignment, while the overall embedding distribution remains well-dispersed, ensuring robustness and mitigating collapse.

\textbf{Independent Prior Decomposition via Angular Reparameterization.} Although our input supervision separates structural and functional descriptions using LLM-based decomposition, this alone does not guarantee that the corresponding embeddings remain disentangled. Neural encoders may still learn overlapping or correlated representations, especially when both descriptions are paired with the same protein sequence. To enforce semantic separation in the latent space, we introduce a modified MMD-based objective that explicitly encourages independence between the structural and functional embeddings.

We model the latent representation \( X \in \mathbb{R}^N \) as a composition of two disjoint subspaces: one corresponding to function (\( X_1 \)) and the other to structure (\( X_2 \)). To regularize the geometry of the representation space, we assume that \( X \) lies on the unit hypersphere, i.e., \( \|X\|^2 = 1 \), consistent with the uniformity objective used elsewhere in our framework. This constraint enforces a fixed total norm, allowing us to explicitly control how representational capacity is allocated between semantic components. To achieve a smooth and interpretable trade-off, we introduce an angular parameter \( \phi \in [0, \pi / 2] \), and define \( r_1 = \cos \phi \) and \( r_2 = \sin \phi \), such that \( \|X_1\| = r_1 \) and \( \|X_2\| = r_2 \). This angular reparameterization ensures that the functional and structural subspaces lie on disjoint hyperspheres of radii \( r_1 \) and \( r_2 \), respectively, and provides a principled way to control their relative contributions to the overall representation.

To sample these priors, we generate i.i.d. Gaussian noise for each subspace and normalize them to the specified radii. This yields two independent latent distributions: one for structure and one for function. Unlike traditional autoencoder frameworks that apply a single isotropic prior to the entire latent space, we introduce Angular MMD, a variant tailored to our reparameterized latent space. Specifically, we apply separate MMD terms to match the learned embeddings for function and structure $\left(Z_{\mathrm{f}}, Z_{\mathrm{s}}\right)$ to their respective angular priors $X_1$ and $X_2$ as denoted in Equation~\ref{eq:loss_dis}. This angular formulation introduces independent priors for structure and function, ensuring that their embeddings not only occupy distinct subspaces but are also distributed across disjoint normed regions of the hypersphere. This encourages semantic disentanglement in both direction and magnitude.

\begin{equation}
\label{eq:loss_dis}
\mathcal{L}_{\text{D}} = \text{MMD}(Z_{\text{f}}, X_1) + \text{MMD}(Z_{\text{s}}, X_2).
\end{equation}

Finally, Our full training loss function can be formulated as Equation~\ref{eq:loss_eq}, where $\lambda_U$, and $\lambda_{\text{D}}$ balance the contributions of uniformity and disentanglement. Together, these components enable controllable and interpretable protein editing by selectively modifying either structural or functional inputs at inference time.

\begin{align}
\label{eq:loss_eq}
\mathcal{L}_{E} = \mathcal{L}_{\text{A}} + \lambda_U \mathcal{L}_{\text{U}} + \lambda_{\text{D}} \mathcal{L}_{\text{D}}.
\end{align}

\textbf{Protein Editing.}  As illustrated in Figure~\ref{fig:full-framework}(c), we perform protein editing by modifying the structural text input, the functional text input, or both. The updated protein embedding is computed via spherical linear interpolation (slerp) between the original protein embedding and the new text-derived embedding that reflects the intended edit, as formalized in Equation~\ref{eq:protein_edit}. \( \alpha \in [0, 1] \) is the interpolation factor (we used 0.9), \( m \in \{0,1\}^d \) is a binary mask that controls which subspace (structure or function) is edited, and \( \odot \) represents element-wise multiplication.

\begin{equation}
z^{\text{edit}} = \text{Slerp}\left(z_p, \ z_t, \ \alpha \right) \odot m + z_p \odot (1 - m).
\label{eq:protein_edit}
\end{equation}

The resulting edited embedding blends the properties of the original and modified inputs and is then passed to the decoder to reconstruct the edited protein sequence. Since we have trained the model to partition the latent space such that the first half encodes structural semantics and the second half encodes functional semantics, we can selectively apply edits to only the relevant subspace during inference. If a single attribute is modified, we retain the unedited portion of the original embedding and interpolate only the corresponding half. For example, if only the functional description is updated, we preserve the structural half of the original protein embedding and apply interpolation only to the functional subspace. When both attributes are edited, we interpolate across the full embedding.

\begin{table*}[t]
\centering
\caption{Performance on protein editing for structure and function editing tasks. Metrics reflect successful edit rate (\%) for each category. The signs (+ or -) indicate whether the attribute is instructed to increase or decrease.}
\scalebox{0.9}{
\renewcommand{\arraystretch}{1.2}
\setlength{\tabcolsep}{5pt}
\begin{tabular}{l|cccc|cccc}
\hline
\textbf{Method} & \multicolumn{4}{c|}{\textbf{Structure}} & \multicolumn{4}{c}{\textbf{Functional}} \\
\cline{2-9}
& + $\alpha$-helices  & - $\alpha$-helices & + $\beta$-sheets & - $\beta$-sheets & + Villin & - Villin & + Pin1 & - Pin1 \\
\hline
ProteinDT~\cite{liu2023proteinDT} & 28.27 & \textbf{69.40} & 9.16 & \textbf{82.85} & 1.41 & 98.59 & 6.25 & 93.75 \\
DisProtEdit ($\lambda_U=0.2, \lambda_D=0$) & 35.87 & 61.01 & 27.10 & 67.64 & 0.00 & 100.00 & 1.56 & \textbf{98.44} \\
DisProtEdit ($\lambda_U=0.2, \lambda_D=0.1$) & \textbf{56.14} & 43.86 & 12.87 & 81.48 & 0.00 & 100.00 & \textbf{14.06} & 87.50 \\
DisProtEdit ($\lambda_U=0.2, \lambda_D=0.5$) & 38.60 & 57.89 & 28.27 & 71.54 & 0.00 & 100.00 & 4.69 & 96.88 \\
DisProtEdit ($\lambda_U=0.2, \lambda_D=0.8$) & 43.66 & 54.78 & 16.96 & 76.61 & 2.82 & 97.18 & 1.56 & 96.88 \\
DisProtEdit ($\lambda_U=0.2, \lambda_D=1.0$) & 48.93 & 48.34 & \textbf{31.58} & 68.23 & 0.00 & \textbf{100.00} & 3.12 & 96.88 \\
DisProtEdit ($\lambda_U=0.2, \lambda_D=5.0$) & 51.66 & 49.12 & 31.38 & 67.06 & \textbf{10.94} & 89.06 & 2.82 & \textbf{97.18} \\
\hline
\end{tabular} }
\label{tab:single-editing-results}
\end{table*}

\begin{table*}[h]
\centering
\caption{Both-hit ratios (\%) for all structure-function editing combinations. Metrics reflect successful edit rate (\%) for each category. The signs (+ or -) indicate whether the attribute is instructed to increase or decrease. All DisProtEdit models are trained with $\lambda_U = 0.2$.}
\scalebox{0.99}{
\small
\begin{tabular}{|l|c|c|c|c|c|c|}
\makecell{\textbf{Combination} \\ \textbf{Editing}} & 
\makecell{\textbf{DisProtEdit} \\ $\lambda_D=0$} & \makecell{\textbf{DisProtEdit} \\ $\lambda_D=0.1$} & \makecell{\textbf{DisProtEdit} \\ $\lambda_D=0.5$} & \makecell{\textbf{DisProtEdit} \\ $\lambda_D=0.8$} & \makecell{\textbf{DisProtEdit} \\ $\lambda_D=1.0$} & \makecell{\textbf{DisProtEdit} \\ $\lambda_D=5.0$} \\
\hline
+ $\alpha$-helices, + Pin1        & 5.10 & 4.08 & 5.61 & 2.04 & 5.10 & \textbf{9.69} \\
+ $\alpha$-helices, - Pin1        & 46.94 & \textbf{61.73} & 45.92 & 47.96 & 43.88 & 50.51 \\
- $\alpha$-helices, + Pin1        & 8.67 & 3.06 & 4.59 & 3.06 & 5.61 & \textbf{8.67} \\
- $\alpha$-helices, - Pin1        & 28.57 & 25.00 & 38.27 & \textbf{43.88} & 42.35 & 32.14 \\
+ $\alpha$-helices, + Villin      & 4.08 & 3.57 & 2.04 & 2.55 & 3.57 & \textbf{7.65} \\
+ $\alpha$-helices, - Villin      & 44.90 & \textbf{53.57} & 46.43 & 48.47 & 47.96 & 50.00 \\
- $\alpha$-helices, + Villin      & 7.65 & 3.57 & 6.12 & \textbf{6.63} & 5.61 & \textbf{6.63} \\
- $\alpha$-helices, - Villin      & 30.10 & 28.57 & 41.33 & 41.84 & \textbf{43.37} & 33.16 \\
+ $\beta$-sheets, + Pin1         & 3.57 & 0.51 & 2.55 & 1.53 & 3.06 & \textbf{5.10} \\
+ $\beta$-sheets, - Pin1         & 12.24 & 3.57 & \textbf{34.69} & \textbf{34.69} & 22.45 & 28.06 \\
- $\beta$-sheets, + Pin1         & 7.65 & 2.04 & 6.12 & 6.63 & 4.08 & \textbf{10.71} \\
- $\beta$-sheets, - Pin1         & 43.88 & \textbf{59.18} & 39.80 & 39.29 & 52.04 & 39.29 \\
+ $\beta$-sheets, + Villin       & 3.57 & 1.53 & 1.53 & 0.51 & \textbf{3.57} & 2.04 \\
+ $\beta$-sheets, - Villin       & 13.27 & 5.61 & \textbf{29.59} & 26.53 & 21.94 & 26.02 \\
- $\beta$-sheets, + Villin       & 7.65 & 4.08 & 7.65 & 6.63 & \textbf{8.16} & 7.14 \\
- $\beta$-sheets, - Villin       & 43.88 & \textbf{59.18} & 42.35 & 47.96 & 52.04 & 47.45 \\
\end{tabular}}
\label{tab:multi-editing-results}
\end{table*}

\subsection{SwissProtDis Dataset}

We construct a dataset comprising approximately 540,000 protein–text pairs, where each protein sequence is associated with a descriptive annotation. The dataset preparation pipeline is illustrated in Figure~\ref{fig:full-framework}(d). Starting from an existing protein–text pair dataset SwissProt~\cite{uniprot}, we employ a large language model (GPT-4o)~\cite{openai2023gpt4} to automatically decompose each raw annotation into two distinct components: (1) a structural description, capturing physical or biochemical properties (e.g., secondary structure, localization), and (2) a functional description, reflecting the protein’s biological role or activity (e.g., catalytic function, signaling behavior). This augmentation process transforms each protein entry into a triplet consisting of the amino acid sequence, a structural text, and a functional text, enabling the support on disentangled representation learning. The resulting dataset, which we refer to as \textbf{SwissProtDis}, serves as the foundation for our disentanglement-based training and enables targeted editing via text modification. We found GPT-4o yield the most reasonable outputs compared to Gemini-1.5-Pro~\cite{geminiteam2024gemini} and GPT3.5~\cite{openai2023gpt4}. The sample entries and the instruction prompt can be found in Appendix~\ref{app:dataset}. 

\section{Experiments}

\subsection{Implementation Details}

Each text is independently encoded using SciBERT~\cite{Beltagy2019SciBERT}, followed by a modality-specific multilayer perceptron (MLP) that projects the output into a shared latent space. Together, the SciBERT encoder and the corresponding MLP form the dual-channel text encoders. For protein sequences, we use ProtBERT~\cite{Elnaggar2020.07.12.199554} to encode amino acid sequences, followed by an MLP projection layer to align the protein embedding with the dimensionality of the text embeddings. This enables cross-modal comparison and latent interpolation. For the disentanglement loss \( \mathcal{L}_{\text{D}} \), we set the angular decomposition parameters to \( r_1^2 = 0.5 \) and \( r_2^2 = 0.5 \), ensuring an equal split between structural and functional components. We find that setting \( \lambda_U = 0.2 \) and \( \lambda_{\text{D}} = 1.0 \) yields the most stable and effective training, as increasing \( \lambda_U \) beyond 0.2 often leads to unstable optimization and \( \lambda_D = 1.0 \) achieves better disentanglement. To reconstruct protein sequences from edited embeddings, we train a autoregressive T5 decoder~\cite{raffel2023exploringlimitstransferlearning} conditioned on the learned representations. All models are fine-tuned from pretrained checkpoints. More implementation details can be found in Appendix~\ref{app:training},~\ref{app:editing}, and~\ref{app:tape}.

\begin{table*}[h]
\centering
\caption{Performance on the TAPE benchmark~\cite{tape2019} across structure prediction, homology classification, and regression tasks. Results for classification tasks (SS-Q3, SS-Q8, Homology) are reported as accuracy, while regression tasks (Fluorescence, Stability) are reported as Spearman's correlation.}
\scalebox{0.9}{
\renewcommand{\arraystretch}{1.2}
\setlength{\tabcolsep}{6pt}
\begin{tabular}{l|c|c|c|c|c}
\hline
\textbf{Method} & \textbf{SS-Q3} & \textbf{SS-Q8} & \textbf{Homology} & \textbf{Fluorescence} & \textbf{Stability} \\
\hline
ProtBert-BFD~\cite{Elnaggar2020.07.12.199554} & 0.8290 & 0.6818 & 0.2381 & 0.3453 & 0.8021 \\
OntoProtein~\cite{zhang2022ontoprotein} & 0.8181 & 0.6758 & 0.2716 & -0.0832 & 0.7110 \\
ProteinDT-InfoNCE~\cite{liu2023proteinDT} & \textbf{0.8329} & \textbf{0.6925} & \textbf{0.3147} & -0.0762 & 0.7356 \\
ProteinDT-EBM-NCE~\cite{liu2023proteinDT} & 0.8326 & 0.6913 & 0.2855 & 0.0167 & 0.7952 \\
DisProtEdit ($\lambda_U=0.2, \lambda_D=0$) & 0.8272 & 0.6577 & 0.2924 & 0.2576 & 0.7731 \\
DisProtEdit ($\lambda_U=0.2, \lambda_D=0.1$) & 0.8287 & 0.6765 & 0.3064 & 0.2760 & 0.8089 \\
DisProtEdit ($\lambda_U=0.2, \lambda_D=0.5$) & 0.8278 & 0.6757 & 0.3022 & 0.1614 & 0.7886 \\
DisProtEdit ($\lambda_U=0.2, \lambda_D=0.8$) & 0.8287 & 0.6763 & 0.3050 & 0.5123 & 0.7897 \\
DisProtEdit ($\lambda_U=0.2, \lambda_D=1.0$) & 0.8285 & 0.6754 & 0.3133 & \textbf{0.5373} & \textbf{0.8258} \\
DisProtEdit ($\lambda_U=0.2, \lambda_D=5.0$) & 0.8276 & 0.6745 & 0.3092 & 0.2503 & 0.8149 \\
\hline
\end{tabular} }
\label{tab:tape-results}
\end{table*}
\subsection{Protein Editing Evaluations}

We evaluate DisProtEdit’s ability to perform controllable protein editing through latent interpolation guided by textual modifications. We consider two editing scenarios: (1) single-attribute editing, where either the structural or functional input is modified while the other remains unchanged, and (2) multi-attribute editing, where both structural and functional descriptions are edited simultaneously. Editing tasks are further categorized into structural edits, which modify secondary structure features such as alpha-helices or beta-sheets, and functional edits, which target protein-specific stability, including Villin and Pin1. For single-attribute editing, we follow the benchmark protocol from prior work~\cite{liu2023proteinDT}. For multi-attribute editing, since no prior work has explored this setting, we construct a new evaluation set by bootstrapping 196 protein sequences and applying paired structure–function edit instructions to them. To quantify editing performance, we use pretrained oracle predictors to assess whether the edited sequence satisfies the intended attribute change(s). Let \( Q^{\text{orig}}_{i,k} \) and \( Q^{\text{edit}}_{i,k} \) denote the predicted property scores for attribute \( k \) of sample \( i \), and let \( \delta_{i,k} \in \{-1, +1\} \) indicate the intended direction of change (decrease or increase, respectively). We define editing accuracy in Equation~\ref{eq:edit-accuracy}, where \( \mathbb{I}[\cdot] \) is the indicator function. A sample is counted as correct only if all its targeted attributes are successfully edited in the intended directions.

\begin{equation}
\text{Acc} = \frac{1}{N} \sum_{i=1}^{N} \mathbb{I} \left[ \bigwedge_{k=1}^{K} \left( \delta_{i,k} \cdot \left( Q^{\text{edit}}_{i,k} - Q^{\text{orig}}_{i,k} \right) > 0 \right) \right].
\label{eq:edit-accuracy}
\end{equation}

In Table~\ref{tab:single-editing-results}, we evaluate DisProtEdit on isolated structural and functional protein editing tasks. For structure editing, DisProtEdit achieves a success rate of 56.14\% in the $+\alpha$-helice condition and 31.58\% in $+\beta$-sheet under $\lambda_D=0.1$ and $\lambda_D=1.0$ respectively, both substantially outperforming ProteinDT (28.27\% and 9.16\%). In contrast, ProteinDT performs best on reduction tasks, but these numbers likely reflect generic destabilization rather than controllable editing. DisProtEdit, on the other hand, provides more controllable behavior. For example, although its $-\alpha$-helice and $-\beta$-sheets success rates vary depending on $\lambda_D$, the model maintains over 65\% accuracy on $-\beta$-sheets across most settings. These results suggest DisProtEdit achieves more targeted structure modulation rather than generic degradation. For function editing, functional improvement tasks remain challenging. $+$Villin reaches only 10.94\% at best ($\lambda_D=5.0$), and $+$Pin1 tops at 14.06\% ($\lambda_D=0.1$). This asymmetry suggests that reducing protein stability is easier than enhancing it. likely due to the ruggedness of the protein fitness landscape and the higher tolerance for disruptive mutations. 

\begin{figure}[t]
    \centering
    \includegraphics[width=0.9\linewidth]{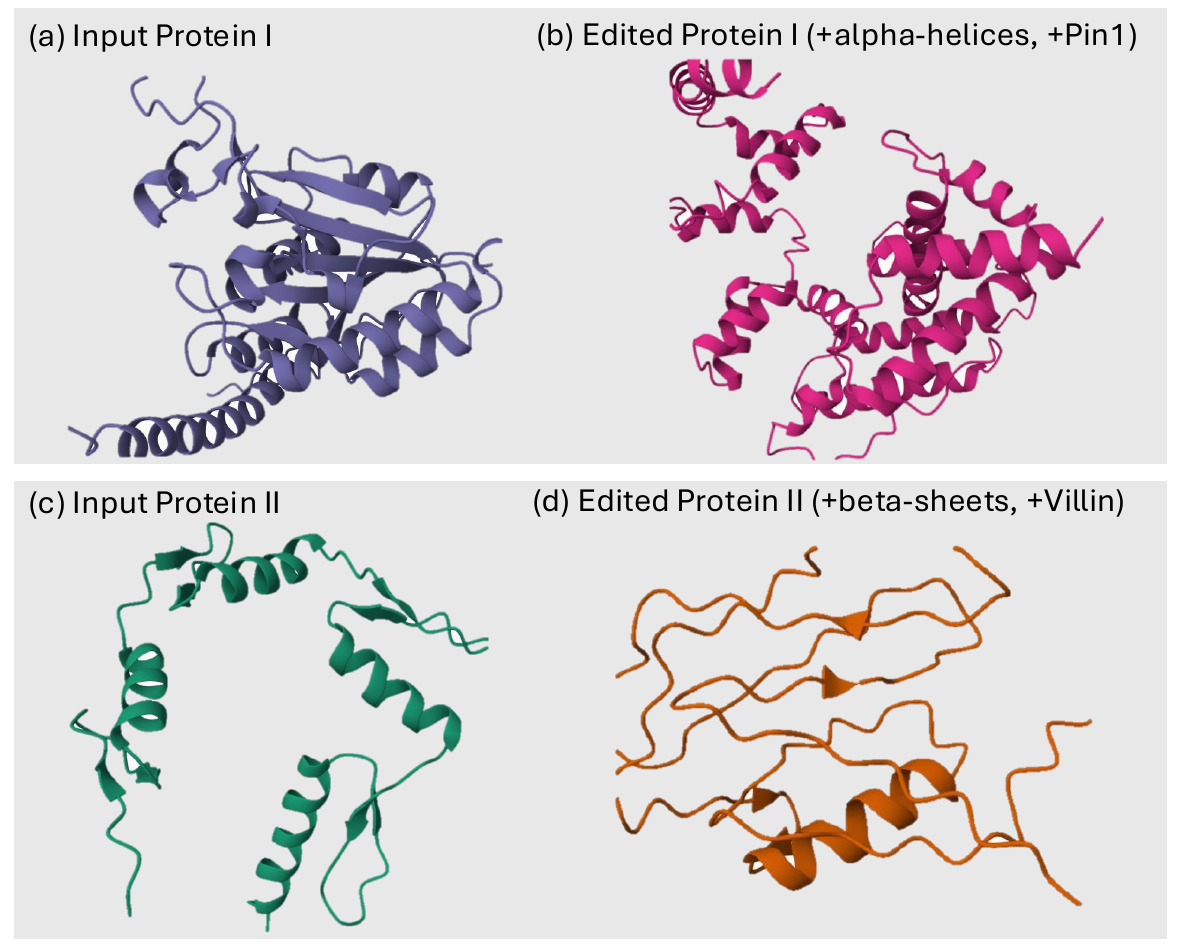}
    \vspace{-1em}
    \caption{
    \textbf{Qualitative visualization of structure–function protein edit samples.}  
    (a, c) Original protein sequences with their corresponding structural and functional attributes.  
    (b, d) Edited proteins generated by DisProtEdit in response to compositional prompts: (b) increase alpha-helices and increase Pin1 stability; (d) increase beta-sheets and increase Villin stability.  
    We showcase these examples because they represent some of the most challenging edit combinations in our benchmark, where the model exhibited the lowest success rates. Despite the difficulty, DisProtEdit demonstrates the ability to generate meaningful multi-attribute modifications in both structure and function.
    }
    \label{fig:edit_visualization}
\end{figure}

In Table~\ref{tab:multi-editing-results}, we analyze how varying the disentanglement weight $\lambda_D$ affects DisProtEdit's performance on multi-attribute editing, where both structure and function are modified simultaneously. We observe that moderate $\lambda_D$ values (0.1 to 1.0) generally achieve the best balance between editing success and disentangled control. Specifically, $\lambda_D = 0.1$ achieves the highest both-hit success in several compatible directions, such as $+\alpha$-helice, $-$Pin1 (61.73\%) and $-\beta$-sheets, $-$Villin (59.18\%). To further illustrate these challenging cases, We visualize the edited protein structures using AlphaFold2 with MMseqs2 ~\cite{ColabFold_Mirdita2022, AlphaFold2021} in Figure~\ref{fig:edit_visualization}, showing qualitative examples of low-success edit combinations. Despite their difficulty, the model still produces plausible edits, demonstrating its capacity for multi-attribute modulation. Moreover, excessively high $\lambda_D$ (e.g., 5.0) sometimes boosts rare cases like $+$Villin but often harms  the successful edit rate in harder tasks, suggesting a trade-off between disentanglement strength and editability.

\subsection{Protein Property Prediction}

To assess the quality and generalizability of our learned protein representations, we evaluate DisProtEdit on four tasks from the TAPE benchmark~\cite{tape2019}: secondary structure prediction (SS-3 and SS-8), remote homology detection, fluorescence prediction, and stability prediction. The first two are classification tasks, where we fine-tune a linear classifier on pooled embeddings and report per-residue accuracy (SS) or fold-level accuracy (homology). The latter two are regression tasks, for which we apply a single-layer MLP and report Spearman’s rank correlation. We omit the contact prediction task from TAPE, as it is not directly aligned with our objective of learning global, semantically disentangled protein representations. All evaluations follow standard TAPE protocols~\cite{tape2019, liu2023proteinDT, zhang2022ontoprotein}.

In Table~\ref{tab:tape-results}, we evaluate the quality of learned protein embeddings and found that DisProtEdit achieves strong performance across both classification and regression tasks. Our best model (\( \lambda_U = 0.2, \lambda_D = 1.0 \)) attains 82.9\% accuracy on secondary structure prediction (SS-Q3), 67.5\% on SS-Q8, and 0.3133 accuracy on remote homology classification, which is comparable to strong baselines such as ProteinDT. On the regression tasks, DisProtEdit achieves a fluorescence correlation of 0.5373 and a stability correlation of 0.8258, outperforming all baselines in both metrics. We also observe that the reproduced fluorescence scores for baselines such as ProteinDT are substantially lower than originally reported, suggesting inconsistencies in evaluation setups. These results demonstrate that the representations learned by DisProtEdit are not only controllable for editing but also competitive for downstream tasks.

\subsection{Discussions}

\begin{figure*}[!t]
    \centering
    \includegraphics[width=0.82\linewidth]{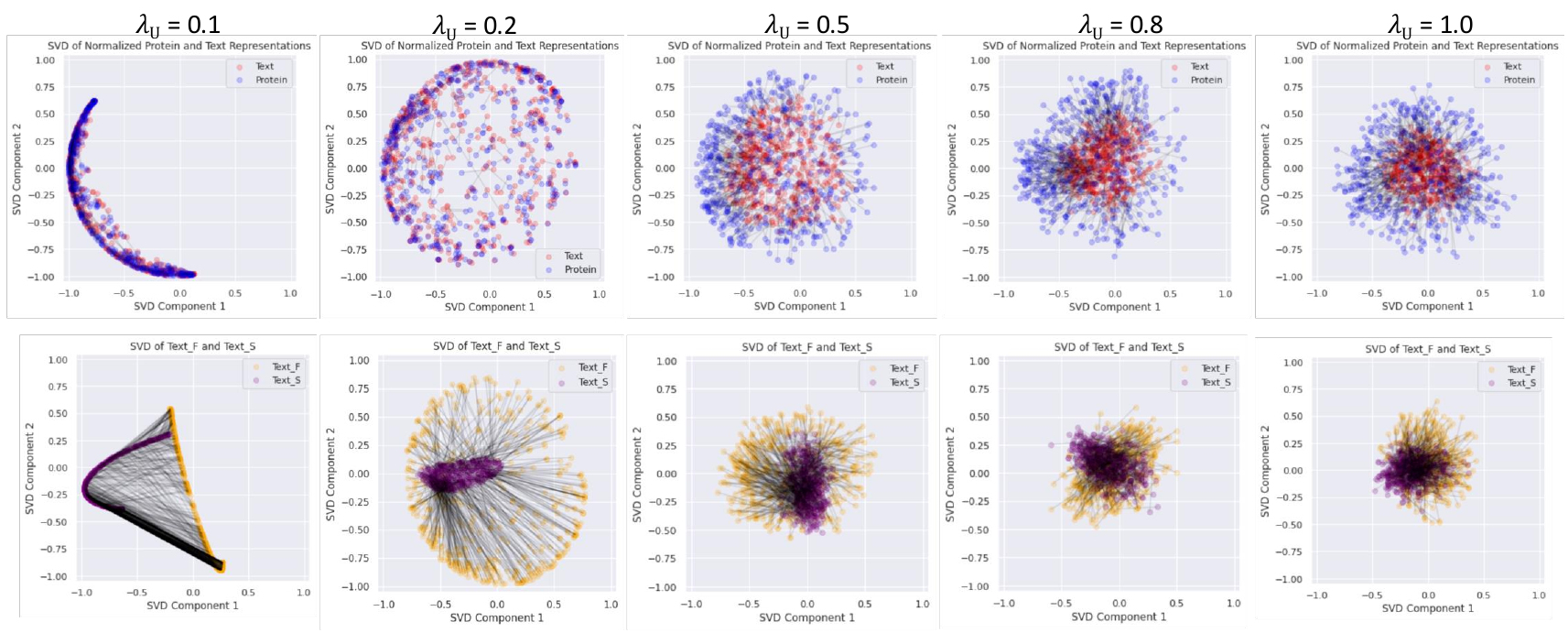}
    \vspace{-1em}
    \caption{Effect of the uniformity loss weight $\lambda_U$ on cross-modal alignment. The plots compare embedding distributions of functional text and structural text representations under varying values.}
    \label{fig:ablation_uniform}
\end{figure*}

\begin{figure*}[!t]
    \centering
    \includegraphics[width=0.82\linewidth]{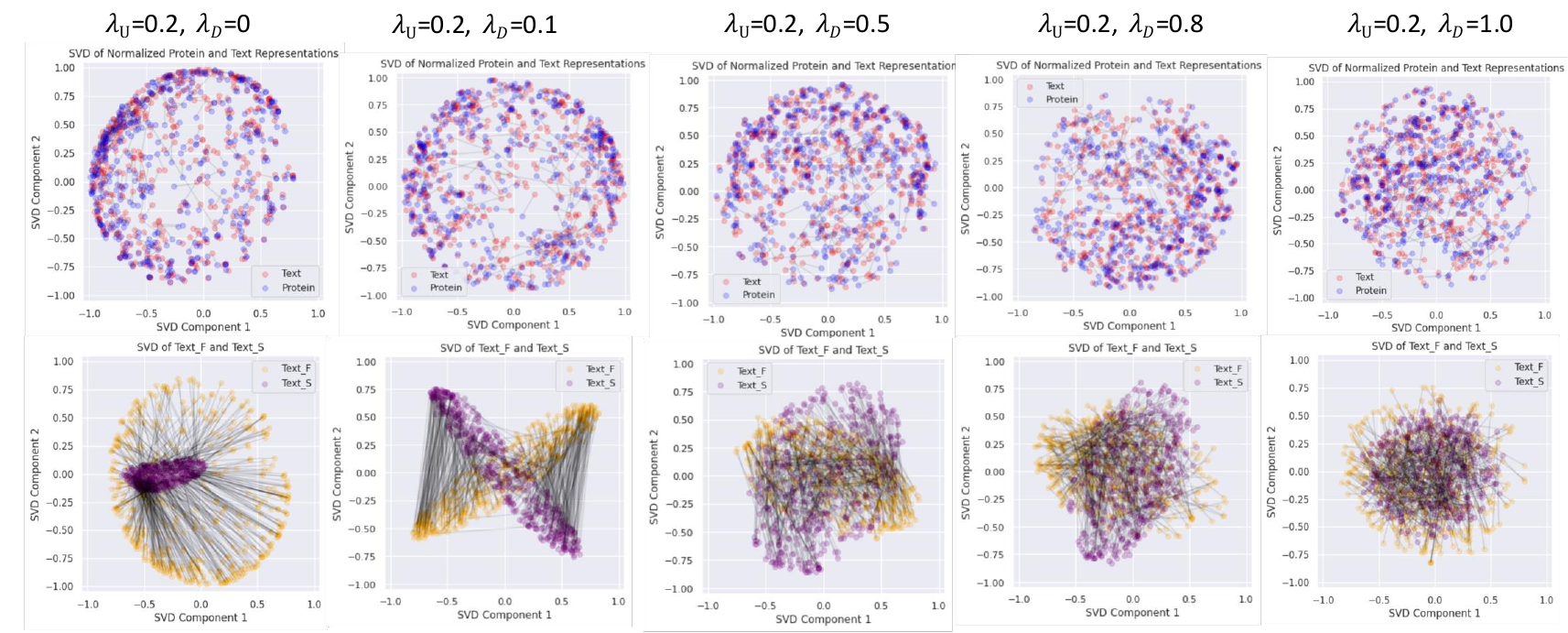}
    \vspace{-1em}
    \caption{Effect of the disentanglement loss weight $\lambda_D$ on cross-modal alignment, comparing embedding distributions of functional text and structural text representations under varying values.
 }
    \label{fig:ablation_dis}
\end{figure*}

\textbf{Visualizing the Representations under Different Training Strategies.} In Figure~\ref{fig:constrastive_comp}, we visualize UMAP projections~\cite{mcinnes2018umap-software} of protein and text embeddings learned under different training strategies: (a) random projection, (b) contrastive learning, (c) contrastive learning followed by fine-tuning with alignment loss, and (d) DisProtEdit (ours). While contrastive learning improves cross-modal alignment over random projection, it still leaves a noticeable modality gap. This gap can be reduced with an additional fine-tuning stage using alignment loss, as shown in (c). In contrast, DisProtEdit achieves comparable cross-modal integration and semantic disentanglement in a single-stage training setup. This demonstrates the effectiveness of our alignment and uniformity objectives for learning meaningful, multimodal representations without requiring separate post-hoc alignment.

\textbf{Visualizing the Representations with Different Loss Weights.}  
We conduct an ablation study on the loss weights \( \lambda_U \) (uniformity) and \( \lambda_D \) (disentanglement) to examine their effects on representation geometry and training dynamics. As shown in Figure~\ref{fig:ablation_uniform}, which presents SVD projections of the learned embeddings, we observe that small values of \( \lambda_U \) (e.g., 0.1) result in a curved, ``banana-shaped'' embedding distribution, indicating insufficient dispersion. In contrast, setting \( \lambda_U = 0.2 \) produces a well-formed spherical embedding structure that promotes diversity while maintaining alignment. However, increasing \( \lambda_U \) beyond 0.2 leads to unstable training, often causing gradient explosion and severe modality misalignment. Figure~\ref{fig:ablation_dis} illustrates the effect of varying the disentanglement loss weight \( \lambda_D \) on the geometry of structural and functional text embeddings. When \( \lambda_D = 0 \), the two modalities are highly entangled, with overlapping distributions that suggest poor semantic separation. As \( \lambda_D \) increases, we observe a gradual divergence between the two subspaces. Notably, at \( \lambda_D = 1.0 \), the separation is most effective. structural text and functional text embedding form two symmetric and coherent clusters. This suggests that the MMD loss at this setting strikes an ideal balance: it encourages semantic independence between structural and functional embeddings without disrupting alignment with the shared protein space. These results confirm that strong but not excessive disentanglement promotes modular representation learning in our context.

\section{Limitations}

\textbf{Quality of LLM-Derived Descriptions and Functional Validation.}  
SwissProtDis relies on a large language model (LLM) to decompose UniProt annotations~\cite{uniprot} into separate structural and functional descriptions. While grounded in expert-curated data, the LLM may introduce hallucinated or biologically imprecise content during decomposition, potentially introducing noise into training and affecting the reliability of learned representations. Furthermore, although our model is designed to disentangle structural and functional factors, we do not explicitly evaluate whether structural edits preserve function or vice versa. A more thorough investigation of these interactions is an important direction for validating the biological plausibility and reliability of the learned representations.

\textbf{Decoder Bias and Reconstruction Inaccuracy.}  
Although the T5 decoder enables reconstruction from latent embeddings, we observe a tendency to overfit to protein fragments that frequently occur in the training data. This results in reduced sequence diversity and occasional inaccuracies, particularly when editing underrepresented or rare motifs. We attribute this to both the limited sequence variability in SwissProtDis—which encourages memorization of common subsequences—and the constrained expressiveness of the decoder architecture. It remains an open question whether these limitations arise primarily from the decoder or the disentangled latent space. Exploring more expressive generative models like diffusion models may improve diversity, robustness, and overall decoding quality.

\textbf{Evaluation and Baseline Limitations.} Evaluating protein editing remains challenging due to the lack of ground-truth labels for most attribute modifications. We rely on pretrained oracle predictors to assess whether edits achieve the intended effect, but these may be noisy or biased, especially for out-of-distribution sequences. Moreover, while we compare DisProtEdit to ProteinDT as a contrastive learning baseline, editing-specific methods such as ProtTex~\cite{ma2025prottexstructureincontextreasoningediting} offer complementary approaches but lack publicly available code or models. This limits direct comparison. We leave a more comprehensive baseline study to future work for protein editing continue to emerge.

\section{Related Works}

\textbf{Protein Representation Learning.} Recent work has increasingly explored the intersection of protein representation learning and natural language understanding. Early models such as ProGen~\cite{madani2020progenlanguagemodelingprotein} and ProGen2~\cite{nijkamp2023progen} treat protein sequences as a form of language, using autoregressive modeling to generate biologically plausible sequences with controllable functions. Similarly, models like ESM~\cite{esmrives2019biological} and ProtBERT~\cite{10.1093/bioinformatics/btac020} apply masked language modeling to capture sequence semantics and generalize across downstream tasks. ProteinDT~\cite{liu2023proteinDT} uses contrastive learning to align protein and text embeddings, enabling text-guided editing. Pinal~\cite{Dai2024.08.01.606258} introduces a two-stage pipeline that predicts structures from text and then sequences from structures, allowing for controllable generation via language. ProLLaMA~\cite{lv2024prollama} adapts instruction-tuned LLMs for unified protein understanding and generation through multi-task training. Other approaches such as Chroma~\cite{Chroma2023} and AlphaFold~\cite{AlphaFold2021} jointly model sequence and structure, with Chroma leveraging diffusion models and AlphaFold using attention-based structural inference. Despite these advances, none of the above works explore alignment and uniformity objectives in biological representation learning. In this work, we demonstrate that applying these objectives to protein–language embeddings yields better representations and competitive downstream performance.

\textbf{Protein Editing and Disentanglement.} Protein editing in our context refers to the modification of biological sequences with natural language. While recent approaches have enabled conditional generation, disentangled and controllable editing remains underexplored. TCR-dWAE~\cite{li2023disentangledwassersteinautoencodertcell} leveraged a disentangled Wasserstein autoencoder, but the application is only limited to T-cell receptors and the generalization to protein domain remain unexplored. ProtET~\cite{yin2024multimodalclipinformedproteinediting} introduces a multimodal transformer with structure-in-context reasoning, allowing interactive protein editing guided by natural language. ProtTex~\cite{ma2025prottexstructureincontextreasoningediting} combine CLIP-style contrastive alignment with instruction-conditioned generation for text-driven editing. To the best of our knowledge, we are the first to investigate protein editing in a disentangled context to support both functional and structural editing.

\section{Conclusion}

We presented DisProtEdit, a framework for protein editing that learns disentangled structural and functional representations from dual-channel natural language descriptions. By aligning protein sequences with modular text supervision, our method enables interpretable and controllable editing with minimal attribute interference. DisProtEdit achieves competitive performance on both editing and representation learning benchmarks, offering fine-grained control through partial textual modifications. The accompanying SwissProtDis dataset, generated using large language models, provides scalable and high-quality supervision for semantic protein understanding. This work lays the foundation for biologically grounded protein design, with future directions including region-specific editing, full-length protein generation, and structure-aware decoding.

\clearpage
\section*{Impact Statement}

We are releasing the SwissProtDis dataset and the multiple attributes editing benchmar on Huggingface dataset with MIT licence. The SwissProtDis dataset contains 540,000 pairs of protein sequence, structural and functional text descriptions. The multiple attributes editing benchmark features 196 samples of protein sequences for multiple attributes editing.

DisProtEdit enables controllable protein editing via text-guided latent manipulation, which may raise dual-use concerns in synthetic biology if applied without proper safeguards. The model relies on oracle predictors and LLM-derived annotations, which can introduce biases or inaccuracies, potentially leading to non-functional or misleading outputs. Additionally, the lack of structural constraints may result in sequences that do not fold correctly. We recommend responsible use alongside expert validation and alignment with biosafety guidelines.


\bibliography{example_paper}
\bibliographystyle{icml2025}

\newpage
\appendix
\onecolumn

\section{Implementation Details of Training}
\label{app:training}

\textbf{Encoders Training Setup.} The encoders training setup is illustrated in Figure~\ref{fig:full-framework}(a). Both the protein and text encoders were optimized with Adam~\cite{kingma2017adammethodstochasticoptimization} using a learning rate of \(1 \times 10^{-5}\) and no learning rate scaling. The model was trained for 10 epochs with a batch size of 24. We applied both alignment and uniformity, and angular MMD objectives during training, using the SwissProtDis dataset with dual-channel supervision.

\textbf{Protein Decoder Training Setup.} The protein decoder training setup is illustrated in Figure~\ref{fig:full-framework}(b). We trained the protein sequence decoder to reconstruct amino acid sequences from the edited latent representations. The decoder architecture was based on a T5 decoder~\cite{raffel2023exploringlimitstransferlearning} model, initialized from pretrained weights. Training was performed using a batch size of 8, a learning rate of \(1 \times 10^{-4}\), and 10 epochs. We used the Adam optimizer for optimization, follow the practice of ProteinDT~\cite{liu2023proteinDT} in training their decoder for protein reconstruction.

\section{Implementation Details of Protein Editing Evaluation}
\label{app:editing}

Table~\ref{tab:edit_prompt} lists the textual prompts used for structural and functional editing in DisProtEdit. Each prompt modifies a single attribute, either secondary structure (e.g., alpha-helix or beta-sheet content) or functional stability (e.g., Villin or Pin1). These prompts are paired with corresponding protein sequences and used during to test the model’s ability to apply edits. Table~\ref{tab:edit_prompt_combo} listed the prompts used in multi-attribute editing benchmark.

\begin{table}[htbp]
\centering
\caption{Prompts for structure editing (S) and functional editing (F) in single-attribute benchmark}
\begin{tabular}{ll}
\toprule
Task & Prompt \\
\midrule
\textbf{+ Alpha Helices} & (S) The amino acid sequence have more alpha helices in the secondary structure. \\
\textbf{- Alpha Helices} & (S) The amino acid sequence have fewer alpha helices in the secondary structure. \\
\textbf{+ Beta Sheets} & (S) The amino acid sequence have more beta sheets in the secondary structure. \\
\textbf{- Beta Sheets} & (S) The amino acid sequence have fewer beta sheets in the secondary structure. \\
\textbf{+ Villin} & (F) The amino acid sequence have higher Villin stability. \\
\textbf{- Villin} & (F) The amino acid sequence have lower Villin stability. \\
\textbf{+ Pin1} & (F) The amino acid sequence have higher Pin1 stability. \\
\textbf{- Pin1} & (F) The amino acid sequence have lower Pin1 stability. \\
\bottomrule
\label{tab:edit_prompt}
\end{tabular}
\end{table}

\begin{table}[htbp]
\centering
\caption{Prompts for combined structure (S) and function (F) editing tasks in multi-attribute editing benchmark}
\begin{tabular}{ll}
\toprule
Task Combination & Prompt \\
\midrule
\textbf{+ Alpha Helices, + Villin} & 
(S) The amino acid sequence has more alpha helices in the secondary structure. \\
& (F) The amino acid sequence has higher Villin stability. \\
\textbf{- Alpha Helices, - Villin} & 
(S) The amino acid sequence has fewer alpha helices in the secondary structure. \\
& (F) The amino acid sequence has lower Villin stability. \\
\textbf{+ Beta Sheets, + Pin1} & 
(S) The amino acid sequence has more beta sheets in the secondary structure. \\
& (F) The amino acid sequence has higher Pin1 stability. \\
\textbf{- Beta Sheets, - Pin1} & 
(S) The amino acid sequence has fewer beta sheets in the secondary structure. \\
& (F) The amino acid sequence has lower Pin1 stability. \\
\textbf{+ Alpha Helices, - Pin1} & 
(S) The amino acid sequence has more alpha helices in the secondary structure. \\
& (F) The amino acid sequence has lower Pin1 stability. \\
\textbf{- Alpha Helices, + Villin} & 
(S) The amino acid sequence has fewer alpha helices in the secondary structure. \\
& (F) The amino acid sequence has higher Villin stability. \\
\bottomrule
\label{tab:edit_prompt_combo}
\end{tabular}
\end{table}

\section{Implementation Details of Protein Property Prediction Evaluation}
\label{app:tape}

To evaluate the quality and generalizability of our learned protein representations, we conduct experiments on selected tasks from the TAPE benchmark. Specifically, we focus on four tasks that span both classification and regression objectives. When finetuned for downstreaming task, we used batch size of 8 and learning rate of $3 \times 10^{-5}$, with 5 epochs, following the standard practice~\cite{tape2019}.

\textbf{Secondary Structure Prediction.} A sequence tagging task where each amino acid in the sequence is assigned a secondary structure label. SS-3 uses a coarse-grained label set (helix, strand, or other), while SS-8 provides finer-grained distinctions. We evaluate accuracy on a per-residue basis, using the standard CB513 test set for consistency with prior works~\cite{tape2019, liu2023proteinDT, zhang2022ontoprotein}.
    
\textbf{Remote Homology Detection.} A sequence classification task where models predict the protein fold family, even under low sequence similarity. Performance is measured using classification accuracy on a held-out set of fold-level labels, assessing the model’s ability to generalize across evolutionary gaps.
    
\textbf{Fluorescence Prediction.} A regression task that models the log-fluorescence intensity of protein variants derived from green fluorescent protein (GFP). Since fluorescence varies continuously, we adopt Spearman’s rank correlation as the evaluation metric, which captures monotonic relationships while being robust to scaling.
    
\textbf{Stability Prediction.} This task involves predicting the thermostability of mutated protein variants. Like fluorescence, stability is evaluated as a continuous property, and we use Spearman’s correlation to quantify prediction quality.

\section{Implementation Detail of Dataset Generation}
\label{app:dataset}

To construct SwissProtDis, we used a large language model (GPT-4o) to decompose existing UniProt annotations into separate structural and functional descriptions. The LLM was prompted with a task-specific instruction (shown in Box) to ensure non-overlapping, interpretable supervision across semantic channels. Table~\ref{tab:example} shows the example entries in SwissProtDis.

\NewTColorBox{Context_Box}{ s O{!htbp} }{%
  floatplacement={#2},
  IfBooleanTF={#1}{float*,width=\textwidth}{float},
  colframe=gray!50!black,colback=gray!10!white,title=Instruction to create SwissProtDis from SwissProt,
  }

\begin{Context_Box}[!ht]
You are a biology expert. Given a FASTA protein sequence and corresponding text description, analyze and provide separate detailed descriptions of the structural and functional properties of the protein. Ensure that:\\
(1) The structural and functional descriptions do not overlap in information.\\
(2) Together, they fully represent the protein's characteristics.
\\
\\
In the structural description, include:\\
- The secondary structure composition (e.g., alpha-helical, beta-sheet, loop regions) with an assessment of whether the alpha-helical content is high or low.\\
- Hydrophobic core formation and stability factors.\\
- Structural motifs and conserved domains contributing to its stability.\\
- Predicted electrostatic interactions and flexibility regions.\\
\\
In the functional description, include:\\
- The biochemical role of the protein (e.g., enzyme, receptor, structural protein).\\
- Its active sites, ligand/cofactor binding regions, and potential catalytic function.\\
- Its interactions with other biomolecules, including potential signaling roles.\\
- Predicted cellular localization and its role in physiological processes.\\
\\
protein sequence: [input protein sequence]
\\
text description: [input text description]
\\

Only return the two strings for the structure information and the functional information in json format \{structure: information, functional: information\}
\label{box:prompt}
\end{Context_Box}

\begin{table}[htbp]
\centering
\caption{Examples of text-protein pairs from SwissProtDis Dataset}
\begin{tabular}{>{\raggedright\arraybackslash}p{6.8cm}|>{\raggedright\arraybackslash}p{5cm}|>{\raggedright\arraybackslash}p{5cm}}
\hline
\textbf{Protein Sequence} & \textbf{Structure Description} & \textbf{Functional Description} \\
\hline
\texttt{MVRLFYNPIKYLFYRRSCKKRLRKALKKLNFY HPPKECCQIYRLLENAPGGTYFITENMTNELI MIAKDPVDKKIKSVKLYLTGNYIKINQHYYIN IYMYLMRYNQIYKYPLICFSKYSKIL} &
This protein belongs to the asfivirus MGF 100 family. &
The protein plays a role in virus cell tropism and may be required for efficient virus replication in macrophages. \\
\hline

\texttt{MVRLFHNPIKCLFYRGSRKTREKKLRKSLKKLN FYHPPGDCCQIYRLLENVPGGTYFITENMTNE LIMIVKDSVDKKIKSVKLNFYGSYIKIHQHYYI NIYMYLMRYTQIYKYPLICFNKYSYCNS} &
The protein sequence consists of 107 amino acids, characterized by motifs that are indicative of the asfivirus MGF 100 family. &
Plays a role in virus cell tropism, and may be required for efficient virus replication in macrophages. \\
\hline

\texttt{MVRLFRNPIKCIFYRRSRKIQEKKLRKSLKKLN FYHPPEDCCQIYRLLENVPGGTYFITENMTND LIMVVKDSVDKKIKSIKLYLHGSYIKIHQHYYI NIYMYLMRYTQIYKYPLICFNKYYNI} &
This protein belongs to the asfivirus MGF 100 family, suggesting it shares structural characteristics common to this family. &
The protein plays a role in virus cell tropism and may be required for efficient virus replication in macrophages. \\
\hline

\texttt{MGNKESKYLEMCSEEAWLNIPNIFKCIFIRKL FYNKWLKYQEKKLKKSLKLLSFYHPKKDFVGI RDMLHMAPGGSYFITDNITEEFLMLVVKHPE DGSAEFTKLCLKGSCIVIDGYYYDTLHIFLSE TPDIYKYPLIRYDR} &
The protein is composed of a sequence of 137 amino acids. It belongs to the asfivirus MGF 100 family, which suggests a potential commonality in tertiary or quaternary structural features characteristic of this family. &
This protein plays a role in virus cell tropism and is potentially crucial for efficient virus replication in macrophages. It is expressed during the early phase of the viral replicative cycle, indicating its importance in the initial stages of viral infection. \\
\hline

\texttt{MGNKESKYLEMCSEEAWLNIPNIFKCIFIRKL FYNKWLKYQEKNLEKRLKLLSFYHPKKDFMGI RDMLDMAPGGSYFITDNVTEEFLMLVVKHPE DGSAEFTKLCLKGGCIVIDGFYYDDLHIFITE NPNLYKYPLIHYDR} &
The protein sequence consists of 137 amino acids, with an abundance of lysine (K), leucine (L), and phenylalanine (F) residues, indicating potential structural motifs suitable for protein interactions and stability. It belongs to the asfivirus MGF 100 family, suggesting it may share common structural features with other members of this family. The sequence includes multiple potential phosphorylation sites, and disulfide bonds could form between cysteine (C) residues, possibly contributing to the protein's conformation and stability. &
This protein plays a role in virus cell tropism and may be necessary for efficient virus replication in macrophages, indicating its importance in viral infection processes. It is expressed during the early phase of the viral replicative cycle, suggesting it has a critical role in the initial stages of viral replication. \\
\hline
\label{tab:example}
\end{tabular}
\end{table}


\end{document}